\documentclass[11pt,english]{article}

\usepackage{latexsym}
\usepackage{epsfig}
\usepackage{amssymb}
\usepackage{array}
\usepackage{float}
\usepackage{algorithm}
\usepackage{algorithmic}
\graphicspath{{figures/}{./}}

\newenvironment{proof}{{\bf Proof. } }{{\hfill $\Box$}\vspace{.5pc}}
\newtheorem{theorem}{Theorem}

\newtheorem{definition}[theorem]{Definition}
\newtheorem{lemma}[theorem]{Lemma}

\floatstyle{ruled}

\floatstyle{ruled}
\newfloat{algo}{thp}{lop}
\floatname{algo}{Algorithm}

\newcommand{\IF}[1]{\textbf{if} {#1}}
\newcommand{\THEN}{\textbf{then} }
\newcommand{\ELSE}{\textbf{else} }

\newcommand{\BEGLIST}{\begin{list}{}{\partopsep -3pt \parsep -2pt}}
\newcommand{\ENDLIST}{\end{list}}

\begin{document}
\title{Scatter of Weak Robots}
 
\author{Yoann Dieudonn\'e \qquad Franck Petit\\
LaRIA, CNRS, Universit\'{e} de Picardie Jules Verne\\
Amiens, France}
\date{}
\maketitle

\section{Introduction}

In this paper, 
we first formalize the problem to be solved, i.e.,
the Scatter Problem (SP). We then show that SP cannot be deterministically solved.  Next, we propose 
a randomized algorithm for this problem.  The proposed solution is trivially self-stabilizing.   
We then show how to design a self-stabilizing version of any deterministic solution 
for the Pattern Formation and the Gathering problems. 

In the next section, we describe the model considered in this paper and the formal definition 
of the problem to be solved, i.e., the \emph{Scatter Problem}.  Next, in Section~\ref{sec:algo}, 
we consider how this problem can be solved.  
We first show that the Scatter Problem cannot be deterministically solved in the considered model.  
We then give a probabilistic algorithm for this problem along with its correctness proof. 
In Section~\ref{sec:rel}, we put the result of Section~\ref{sec:algo} back in the context 
of distributed coordination of autonomous mobile robots.  
In this area, two classes of problem received a particular attention\footnote{
	Note that some of the following solutions are in a model called CORDA~\cite{P(1)01} 
	allowing more asynchrony among the robots than the semi-synchronous model (SSM) used 
	in this paper. However, it is showed in~\cite{P(2)01} that any algorithm that 
	correctly solves a problem $P$ in CORDA, correctly solves $P$ in $SSM$.  So,
	any algorithm described in CORDA also works in $SSM$.
}:

\begin{enumerate}
\item The \emph{Pattern Formation Problem} (PFP) which includes 
  the \emph{Circle Formation Problem}, 
 e.g.~\cite{SY99,FPSW99,FPSW01,DK02,K05,DLP06,DP07};
\item The \emph{Gathering Problem} (GP), e.g., \cite{AOSY99,SY99,FPSW99(b),SY99,CP02,CFPS03}.
\end{enumerate}

We consider this two major classes of problems into self-stabilization settings.    
In a self-stabilizing system, regardless of the initial states of the computing units, 
is guaranteed to converge to the intended behavior in finite time~\cite{D74,D00}.  
To our best knowledge, all the above solutions assume that in the initial configuration, 
no two robots are located at the same position.  As already noticed~\cite{DK02,H06}, 
this implies that none of them is self-stabilizing.  
In Section~\ref{sec:rel}, we show that, being self-stabilizing, the proposed algorithm can 
be used to provide a self-stabilizing version of any deterministic solution for PFP and GP, 
i.e., assuming any arbitrary initial configuration---\emph{including} configurations 
where two or more robots can be located at the same position.
Finally, we conclude the story in Section~\ref{sec:conclu}. 

\section{Preliminaries}
\label{sec:model}

In this section, we define the distributed system, basic definitions and the problem considered in this paper.  

\paragraph{Distributed Model.}

We adopt the model introduced~\cite{SY96}, in the remainder referred as~$SSM$---stands for \emph{Semi-Synchronous Model}.  
The \emph{distributed system} considered in this paper consists of $n$ mobile \emph{robots} 
(\emph{entity}, \emph{agent}, or \emph{element}) 
$r_{1}, r_{2},\cdots , r_{n}$---the subscripts $1,\ldots ,n$ are used for notational purpose only.
Each robot $r_{i}$, viewed as a point in the Euclidean plane, moves on this two-dimensional 
space unbounded and devoid of any landmark.  When no ambiguity arises, $r_{i}$ also denotes the 
point in the plane occupied by that robot. 
It is assumed that the robots never collide and that two or more robots may 
simultaneously occupy the same physical location.
Any robot can observe, compute and move with infinite decimal precision.
The robots are equipped with sensors allowing to detect the instantaneous position of the other robots in the plane. 
Each robot has its own local coordinate system and unit measure.  
There is no kind of explicit communication medium. The robots implicitly ``communicate'' by 
observing the position of the others robots in the plane, and by executing a part of their program accordingly. 

The considered robots are \emph{uniform}, \emph{oblivious}, 
and \emph{anonymous}.  The former indicates that
they all follow the same program.  
Obliviousness states that the robots cannot remember any previous 
observation nor computation performed in any previous step.  
Anonymous means that no local parameter 
(such that an identity) which could be used in the program code to differentiate any of them.

In this paper, we also discuss some capabilities the robots are able to have or not:
\BEGLIST
\item [\emph{Multiplicity Detection}:]
The robots are able to distinguish whether there is more than
one robot at a given position;
\item [\emph{Localization Knowledge}:] 
The robots share a common coordinate system, i.e., a common Cartesian coordinate system 
with a common origin and common $x$-$y$ axes with the same orientations. 
\ENDLIST

Time is represented as an infinite sequence of time instant $t_0, t_1, \ldots, t_j, \ldots$ 
Let $P(t_j)$ be the set of the positions in the plane occupied by the $n$ 
robots at time $t_j$ ($j\geq0$). For every $t_j$, $P(t_j)$ is called the \emph{configuration} 
of the distributed system in $t_j$.
$P(t_j)$ expressed in the local coordinate system of any robot $r_i$ is called a \emph{view}, denoted 
$v_i(t_j)$.
At each time instant $t_j$ ($j\geq 0$), each robot $r_i$ is either {\it active} or {\it inactive}. 
The former means that, during the computation step $(t_j,t_{j+1})$, using 
a given algorithm, $r_i$ computes in its local coordinate system a position $p_i(t_{j+1})$ depending 
only on the system configuration at $t_j$, and moves towards $p_i(t_{j+1})$---$p_i(t_{j+1})$ can be equal to
$p_i(t_j)$, making the location of $r_i$ unchanged.
In the latter case, $r_i$ does not perform any local computation and remains at the same position.  In every single activation, the distance 
traveling by any robot $r$ is bounded by $\sigma_r$. So, if the destination point computed by $r$ is farther than $\sigma_r$, then $r$ moves 
toward a point of at most $\sigma_r$. This distance may be different between two robots.

The concurrent activation of robots is modeled by 
the interleaving model in which the robot activations are driven by a \emph{fair scheduler}.
At each instant $t_j$ ($j \geq 0$), the scheduler arbitrarily activates a (non empty) set of robots.   
Fairness means that every robot is infinitely often activated by the scheduler.

\paragraph{Specification.}
The \emph{Scatter Problem} (SP) is to design a protocol for $n$ mobile autonomous robots so that
the following properties are true in every execution:
\BEGLIST
\item [\emph{Convergence}:] Regardless of the initial position of the robots on the plane, no two robots are eventually
located at the same position. 
\item [\emph{Closure}:] Starting from a configuration where no two robots are located at the same position, no two robots
are located at the same position thereafter. 
\ENDLIST

\section{Algorithm}
\label{sec:algo}
The scope of this section is twofold. We first show that,there exists no deterministic algorithm solving $SP$. The result holds even if the 
robots are not oblivious, share a common coordinate system, or are able to detect multiplicity.
Next, we propose a randomized algorithm which converges toward a distribution where the robots have distinct positions. 

\subsection{Impossibility of a Deterministic Algorithm}

\begin{lemma}
\label{lem:imp} 
There exists no deterministic algorithm that solves the Scatter Problem in SSM, even if 
the robots have the localization knowledge or are able to detect the multiplicity.
\end{lemma}
\begin{proof}
Assume, by contradiction, that a deterministic algorithm $A$ exists solving SP in SSM with robots having 
the localization knowledge and being able to detect the multiplicity.  Assume that, initially ($t_0$), 
all the robots are located at the same position.  So, it makes no matter whether the robots have the localization knowledge, are able to detect the multiplicity, or not, 
all the robots have the same view of the world. 
Assume that at $t_0$, all the robots are active and execute $A$.  Since $A$ is a deterministic algorithm and all the robots have the same view, 
then all the robots choose the same behavior.  So, at time $t_1$, all of them share the same position on the place.  Again, they all have the same view of the world.  
By induction, we can deduce that there exists at least one execution of $A$ where the robots always share the same position.  This contradicts the specification of SP.  Hence, such 
an algorithm~$A$ does not exist. 
\end{proof}

Note that Lemma~\ref{lem:imp} also holds whether the robots are oblivious or not.  Indeed, assume non-oblivious robots, i.e., 
any robot moves according to the current and previous configurations.  So, each robot $r_i$ is equipped with a (possibly infinite) 
history register $\mathcal{H}_i$.  At time $t_0$, for each robot $r_i$, the value in $\mathcal{H}_i$ depends on whether the 
registers are assumed to be initialized or not.  

Assume first that, at $t_0$, $\mathcal{H}_i$ is initialized for every robot.  Since the robots are assumed to be uniform and anonymous, 
the values stored in the history registers cannot be different.  So, for every pair of robots $(r_i,r_{i'})$, $\mathcal{H}_i = \mathcal{H}_{i'}$ at $t_0$. Then,
all the robots have the same view of the world.  This case leads to the proof of Lemma~\ref{lem:imp}.  

Now, assume that, for every robot $r_i$, $\mathcal{H}_i$ is not assumed to be initialized at time $t_0$.  Note
that this case captures the concept of self-stabilization. 
In such a system, at $t_0$, one possible initialization of the history registers can be as follows: 
$(r_i,r_{i'})$, $\mathcal{H}_i = \mathcal{H}_{i'}$ for every 
every pair $(r_i,r_{i'})$.  This case is similar to the previous case.  

\subsection{Randomized Algorithm}
\label{sub:algo}

We use the following concept, \emph{Voronoi diagram}, in the design of Algorithm~\ref{algo1}.

\begin{definition}[Voronoi diagram]\cite{A91,DK02}
\label{definition1}
The Voronoi diagram of a set of points $P=\{p_1,p_2,\cdots,p_n\}$ is a subdivision of the plane into $n$ cells, one for each point in $P$. 
The cells have the property that a point $q$ belongs to the Voronoi cell of point $p_i$ iff for any other point $p_j \in P$, 
$dist(q,p_i)< dist(q,p_j)$ where $dist(p,q)$ is the Euclidean distance between $p$ and $q$. In particular, the strict inequality 
means that points located on the boundary of the Voronoi diagram do not belong to any Voronoi cell. 
\end{definition}

We now give an informal description of Procedure~$SP$, shown in Algorithm~\ref{algo1}.  
Each robot uses Function~\emph{Random()}, which 
returns a value randomly and uniformally chosen over $\{0,1\}$. 
When any robot $r_i$ becomes active at time $t_j$, it first computes the Voronoi Diagram of $P_i (t_j)$, i.e., the set of points occupied 
by the robots, $P(t_j)$, computed in its own coordinate system.  Then, $r_i$ moves toward a point inside its Voronoi cell $Cell_i$ whether 
$Random()$ returns $0$. 

\begin{algo}[htb]
\begin{tabbing}
  xxxxx \= xxxxx \= xxxxx \= xxxxx \= xxxxx \= xxxxx \= xxxxx \= xxxxx \= xxxxx \= \kill 
Compute the Voronoi Diagram;\\
$Cell_i$ := the Voronoi cell where $r_i$ is located;\\
$Current\_Pos$ := position where $r_i$ is located;\\ 
\IF{$Random()$=0}\\ 
\THEN \> Move toward an arbitrary position in $Cell_i$, which is different from $Current\_Pos$;\\
\ELSE \> Do not move;
\end{tabbing}
\caption{Procedure $SP$, for any robot $r_i$. \label{algo1}}
\end{algo}

\begin{lemma}[Closure]
\label{lemma1}
For any time $t_j$ and for every pair of robots $(r_i , r_{i'})$ having distinct positions at $t_j$
($p_i(t_j) \neq p_{i'}(t_j)$), then by executing Procedure~$SP$, $r_i$ and $r_{i'}$ remains at distinct positions thereafter 
($\forall j'>j,\ p_i(t_{j'}) \neq p_{i'}(t_{j'})$).
\end{lemma}
\begin{proof}
Clearly, if at time $t_j$, $r_i$ and $r_{i'}$ have distinct positions, then $r_i$ and $r_{i'}$ are in two different Voronoi cells, $V_i$ and $V_j$, respectively. 
 From Definition~\ref{definition1}, $V_i \cap V_j = \emptyset$. 
Furthermore, each robot can move only in its Voronoi cell.  So, we deduce that $r_i$ and $r_{i'}$ have distinct positions at time $t_{j+1}$. 
The lemma follows by induction on $j'$, $j'>j$. 
\end{proof}

In the following, we employ the notation $Pr[A]=v$ to mean that $v$ is the probability that the event $A$ occurs.  
Two events $A$ and $B$ are said to be \emph{mutually exclusive} if and only if $A \cap B = \emptyset$. 
In this case, $Pr[A\cup B]= Pr[A]+Pr[B]$.  The probability that an event $A$ occurs given the known occurrence of 
an event $B$ is the conditional probability of $A$ given $B$, denoted by $Pr[A|B]$.  We have $Pr[A \cap B]=Pr[A|B]Pr[B]$. 

\begin{lemma}[Convergence]
\label{lemma2}
For any time $t_j$ and for every pair of robots $(r_i , r_{i'})$  such that
$p_i(t_j) = p_{i'}(t_j)$. 
By executing Procedure~$SP$, we have  
$$
  \lim_{k\rightarrow\infty} Pr[p_i(t_{j+k}) \neq p_{i'}(t_{j+k})]=1
$$
\end{lemma}

\begin{proof}
Consider at time $t_j$, two robots $r_i$ and $r_{i'}$ such that $p_i(t_j) = p_{i'}(t_j)$. Let $X_{t_j}$ 
(respectively, $Y_{t_j}$) be the 
random variable denoting the number of robots among $r_i$ and $r_{i'}$ which are activated (respectively, move).
$Pr[X_{t_j}=z]$ (resp. $Pr[Y_{t_j}=z']$) indicates the probability that $z \in [0..2]$ (resp. $z' \in [0..2]$) robots among 
$r_i$ and $r_{i'}$ are active (resp.move) at time $t_j$. 
Note that robot $r_i$ (resp $r_{i'}$) can move only if $r_i$ (resp $r_{i'}$) is active. 

Both $r_i$ and $r_{i'}$ are in a single position at time $t_{j+1}$ only if one of the following three events arises 
in the computation step $(t_j,t_{j+1})$:
\begin{itemize}
\item {\bf Event1}: 
  ``Both $r_i$ and $r_{i'}$ are inactive.'' 
  In this case: 
  \begin{equation}\label{eq:1}
    Pr[Event1]=Pr[X_{t_j}=0]\leq1
  \end{equation}  
  
\item {\bf Event2}: 
  ``There is exactly one active robot which does not move and one inactive robot.'' 
  Then, we have:
  $$
    Pr[Event2]=Pr[X_{t_j}=1 \cap Y_{t_j}=0]
  $$
  So,
  $$
    Pr[Event2]=Pr[Y_{t_j}=0|X_{t_j}=1]Pr[X_{t_j}=1]
  $$
  $$
    Pr[Event2]\leq\frac{1}{2}Pr[X_{t_j}=1]
  $$ 
  Thus,
  \begin{equation}\label{eq:2}
    Pr[Event2]\leq\frac{1}{2}
  \end{equation}

\item {\bf Event3:} 
  ``There are exactly two active robots and both of them move toward the same location.''
  The probability that both robots are activated and move (not necessary at the same location) is given by: 
  $$
    Pr[X_{t_j}=2 \cap Y_{t_j}=2]
  $$
 But, 
 $$
   Pr[X_{t_j}=2 \cap Y_{t_j}=2]=Pr[Y_{t_j}=2|X_{t_j}=2]Pr[X_{t_j}=2]
 $$
 That is,
 $$
   Pr[X_{t_j}=2 \cap Y_{t_j}=2]=\frac{1}{4}Pr[X_{t_j}=2]
 $$
 Thus, 
 $$
   Pr[X_{t_j}=2 \cap Y_{t_j}=2]\leq\frac{1}{4}
 $$
 Since the probability that all the robots are activated and move (not necessary at the same location) is lower than or equal to $\frac{1}{4}$,
 the probability of Event3 (i.e both move toward the same location) is also lower than or equal to $\frac{1}{4}$, i.e.
 \begin{equation}\label{eq:3}
 Pr[Event3]\leq\frac{1}{4}
 \end{equation}
\end{itemize}

Let $\Omega$ be a sequence of time instants starting from $t_j$. Denote by $k$ the number of time instants in $\Omega$. 
The value $a$ (resp. $na$) indicates the number of instant in $\Omega$ where at least one robot is active 
(resp. both $r_i$ and $r_{i'}$ are inactive) among $r_i$ and $r_{i'}$.  Obviously, $a+na=k$. 
 From the equations~(\ref{eq:2}) and ~(\ref{eq:3}) and the fact that Event2 and Event3 are mutually exclusive, 
we have: 
$$
  Pr[Event2 \cup Event3]= Pr[Event2]+Pr[Event3]
$$
So, 
\begin{equation}\label{eq:4}
  Pr[Event2 \cup Event3]\leq \frac{1}{2}+\frac{1}{4}=\frac{3}{4}
\end{equation}
 From the equations~(\ref{eq:1}) and~(\ref{eq:4}), the probability that $r_i$ and $r_{i'}$ are located at the same position 
 after $k$ time instant is 
$$
  Pr[p_i(t_{j+k}) = p_{i'}(t_{j+k})]\leq (\frac{3}{4})^aPr[Event1]^{na}\leq(\frac{3}{4})^a
$$ 
By fairness, both $r_i$ and $r_{i'}$ are infinitely often activated. Therefore, $\lim_{k\rightarrow\infty} a = \infty$, and then 
$$
  \lim_{k\rightarrow\infty} Pr[p_i(t_{j+k}) = p_{i'}(t_{j+k})]=0
$$
The lemma follows from the fact that $Pr[p_i(t_{j+k}) \ne p_{i'}(t_{j+k})] = 1-Pr[p_i(t_{j+k}) = p_{i'}(t_{j+k})]$.
\end{proof}

 From Lemma~\ref{lemma1} and~\ref{lemma2} follows:
\begin{theorem}
\label{theorem1}
 Procedure~$SP$ solves the Scatter Problem in SSM with a probability equal to 1.
\end{theorem}

Note that as a result of Theorem~\ref{theorem1} and by the specification of the Scatter Problem,
Procedure~$SP$ provides a self-stabilizing solution in SSM.  

\section{Related Problems and Self-Stabilization}
\label{sec:rel}

The acute reader should have noticed that 
by executing Procedure~$SP$ infinitely often, the robots never stop moving inside their Voronoi cells, even if 
no two robots are located at the same position.  This comes from the fact that Procedure~$SP$ does not require robots 
having the multiplicity detection capability. 
Henceforth in this section, let us assume that the robots are equipped of such an ability.  
This assumption trivially allows the robots to stop if there exists no position with more than 
one robot. 
So, with the multiplicity detection, Procedure~$SP$ provides a valid initial configuration for every solution for PFP and GP.
In the next two subsections, we show how Procedure~$SP$ can be used to provide self-stabilizing algorithms for PFP and GP.

\subsection{Pattern Formation Problem} 

This problem consists in the design of protocols allowing the robots to form 
a specific class of patterns.  

Let Procedure~$A_{PF}(C)$ be a deterministic algorithm in $SSM$ allowing the robots to form a class of pattern $C$. 
Algorithm~\ref{algo2} shows Procedure~$SSA_{PF}(C)$, which can form all the patterns in $C$ starting from any 
arbitrary configuration. 

\begin{algo}[htb]
\begin{tabbing}
  xxxxx \= xxxxx \= xxxxx \= xxxxx \= xxxxx \= xxxxx \= xxxxx \= xxxxx \= xxxxx \= \kill 
\IF{there exists at least \textbf{one} position with a strict multiplicity}\\ 
\THEN \> $SP$;\\
\ELSE \> $A_{PF}$;
\end{tabbing}
\caption{Procedure $SSA_{PF}(C)$ for any robot $r_i$. \label{algo2}}
\end{algo}

\begin{theorem}
\label{theorem2}
 Procedure~$SSA_{PF}(C)$ is a self-stabilizing protocol for the Pattern Formation Problem in SSM 
 with a probability equal to 1.
\end{theorem}

\subsection{Gathering Problem}

This problem consists to make $n\geq2$ robots gathering in a point (not predetermined in advance) in finite time. 
In~\cite{P(2)01}, it has been proved that GP is deterministically unsolvable in $SSM$ and CORDA. 
In fact, one feature that the robots must have in order to solve GP is the multiplicity detection \cite{SY99,CP02,CFPS03}. 
Nevertheless, even with the ability to detect the multiplicity, GP remains unsolvable, in a deterministic way, 
for $n=2$ in $SSM$ \cite{SY99}. For all the other cases ($n\geq3$), GP is solvable. 
So, when $n\geq3$, the common strategy for solving GP is to combine two subproblems which are easier 
to solve. In this way, GP is separated into two distinct steps:

\begin{enumerate}
\item[1.] Starting from an arbitrary configuration wherein all the positions are 
  distinct, the robots must move in such a way to create exactly one position with at least two robots on it;
\item[2.] Then, starting from there, all the robots move toward that unique position with a strict multiplicity.
\end{enumerate}

As for the deterministic algorithms solving PFP, the deterministic algorithm solving GP ($n\geq3$) requires that the 
robots are arbitrarily placed in the plane but with no two robots in the same position. 
Let Procedure~$A_{GP}$ be a deterministic algorithm solving GP, for $n\geq3$, 
with multiplicity detection in $SSM$. Algorithm~\ref{algo4} shows Procedure~$SSA_{GP}$, which solves GP with 
multiplicity detection starting from any arbitrary configuration whether $n\geq3$. 
Remark that it is paradoxical that to make GP self-stabilizing, the robots must scatter before gathering. 

\begin{algo}[htb]
\begin{tabbing}
  xxxxx \= xxxxx \= xxxxx \= xxxxx \= xxxxx \= xxxxx \= xxxxx \= xxxxx \= xxxxx \= \kill 
\IF{there exist at least \textbf{two} positions with a strict multiplicity}\\ 
\THEN \> $SP$;\\
\ELSE \> $A_{GP}$;
\end{tabbing}
\caption{Procedure $SSA_{GP}$ for any robot $r_i$, $n\geq3$. \label{algo4}}
\end{algo}

\begin{theorem}
\label{theorem3}
 Procedure~$SSA_{GP}$ is a self-stabilizing protocol for the Gathering Problem in SSM 
 with a probability equal to 1 whether $n \geq 3$.
\end{theorem}

Note that, for the case $n=2$, we can provide a randomized algorithm solving GP. 
Informally, when any robot becomes active, it chooses to move to the position of the other 
robot with a probability $\frac{1}{2}$.  By using a similar idea as in the proof of Lemma~\ref{lemma2}, 
we can prove that both robots eventually occupy the same position with a probability $1$. 
By combining our basic routine for $n=2$ with Procedure~$SSA_{GP}$, we obtain a procedure 
which solves the self-stabilizing GP with multiplicity detection starting from any arbitrary configuration. 
It follows:

\begin{theorem}
\label{theorem4}
 There exists a self-stabilizing protocol for the Gathering Problem in SSM 
 with a probability equal to 1 for any $n \geq 2$.
\end{theorem}

\section{Conclusion}
\label{sec:conclu}
We shown that the Scatter Problem cannot be deterministically solved.  We proposed 
a randomized self-stabilizing algorithm for this problem. 
We used it to design a self-stabilizing version of any deterministic solution 
for the Pattern Formation and the Gathering problems. 


\end{document}